\documentstyle[aps,prl,multicol,epsf,subfigure]{revtex} 
\def\D{\partial}
\def\grad{\nabla}

\def\eq{\begin{eqnarray}}
\def\qe{\end{eqnarray}}
\def\eqnn{\begin{eqnarray*}}
\def\qenn{\end{eqnarray*}}
\def\nn{\nonumber}
\def\be{\bm{e}}

\def\bq{\bm{q}}
\def\br{\bm{r}}
\def\bu{\bm{u}}

\def\simge{\;\lower3pt\hbox{$\stackrel{\textstyle >}{\sim}$}\;}
\def\simle{\;\lower3pt\hbox{$\stackrel{\textstyle <}{\sim}$}\;}
\def\bm#1{\mbox{\boldmath $#1$}}

\def\lrL#1{\left[#1\right]}

\def\lrS#1{\left(#1\right)}

\def\mycomment#1{}
\def\mycaption#1{\nopagebreak[4]\hspace{0mm}\\\begin{minipage}[h]{85mm}\caption{#1}\end{minipage}\vfill}

\def\f#1#2{\frac{#1}{#2}}
\def\der#1#2{\f{\D #1}{\D #2}}
\def\fder#1#2{\f{\delta #1}{\delta #2}}
\newcommand{\tG}{{\sf G}}

\newcommand{\bR}{\bm{R}}
\title{Pattern Formation in Phase-Separating Gels with Spontaneous Shear}
\author {Nariya Uchida*}
\address{Max-Planck-Institut f\"ur Polymerforschung, 
Postfach 3148, D-55021 Mainz, Germany
}
\date{Submitted August 7, 2001; published as 
Phys. Rev. Lett. {\bf 89}, 025702 (2002)}
\begin{document}
\draft
\bibliographystyle{prsty}
\maketitle
\begin{abstract}
We study pattern formation in gels undergoing
simultaneous phase separation and orientational ordering.
A 2D numerical simulation is performed using
a minimal model of nonlinear elasticity 
with density-anisotropy coupling.
For strong positive coupling, the collapsed phase elongates
along the phase boundary and buckles, creating a folded 
structure with paired topological defects. 
For negative coupling, soft elasticity of the swollen phase
causes a droplet morphology as in liquid-liquid phase separation.
Their possible realizations in 
nematic liquid-crystalline gels are discussed.
\end{abstract}
%
%
\begin{multicols}{2} 
%
%
Polymer gels are unique objects that undergo 
large deformation upon phase transition,
which creates various patterns. 
Characteristic to pattern formation in gels is the 
interplay between rubber-elasticity and interfacial 
tension, as demonstrated by the surface 
undulations seen
in swelling/shrinking processes~\cite{TTanaka}. 
Elasticity also affects phase separation in bulk, which produces 
a foam-like morphology with the polymer-rich region connected~\cite{SSK,OnP}.
A similar morphology is observed in viscoelastic polymer 
solutions~\cite{HTanaka} as well as in polymerization-induced
phase separation~\cite{PDLC}. While these patterns are 
caused by volume phase transition of the polymer network, 
there is a class of amorphous solids in which 
transition induces shear deformation. 
The resulting low-temperature phase is an intermediate 
between usual solid and liquid, in the sense that 
one of the linear shear moduli vanishes in that phase~\cite{GL}. 
We shall refer to it as ``anisotropic glass'' (AG) after Ref.~\cite{GL}. 
An example of AG is provided by nematic 
liquid-crystalline elastomers~\cite{deGennes,Finkelmann}, 
which elongate along the director upon the isotropic-nematic
transition. In their phase ordering process, the spontaneous 
shear deformation creates an anisotropic packing structure 
of orientationally correlated regions, which gives rise to 
an extremely soft nonlinear elastic response~\cite{lcgel1,lcgel2}.
The anisotropy axis of deformation is locally parallel to 
the nematic director, which can be integrated out to give 
an effective free energy as a functional of strain.
The aim of this Letter is to address 
pattern formation in phase-separating AGs,
where volume change and spontaneous shear cooperate~\cite{OM}. 
Motivated by the idea of cell-dynamical systems~\cite{OoP},
we adopt a minimal model of non-linear (visco-)elasticity 
that retains the essential features of (i) fixed-point 
configuration in the order-parameter space and (ii) 
vectorial volume-conserving dynamics. 
%
%
\par
We consider a two-dimensional system for simplicity.
Let $\br$ and $\bR$ be the positions of material element 
before and after deformation, respectively.
The field $\bR(\br)$ defines the 
deformation tensor $\D R_i/\D r_j$ and a symmetric
strain tensor, $G_{ij}=(\D R_k/\D r_i)(\D R_k/\D r_j)$.
For a system with rotational invariance, 
free energy change due to homogeneous 
deformations can be expressed 
in terms of the two invariants
$\sigma = \lrS{\det \tG}^{1/2} = \lambda_1 \lambda_2$
and
$\tau = \mbox{tr\,} \tG/2 = (\lambda_1^2 + \lambda_2^2)/2$,
where $\lambda_1$ and $\lambda_2$ are the
principal elongation ratios.
Note that $\sigma$ is the volume expansion ratio
while $\tau-\sigma=(\lambda_1-\lambda_2)^2/2$
gives a measure of shape anisotropy.
In dimensionless units,
the homogeneous part of our model 
free energy density is written as
\eq
f &=& (\sigma - \sigma_P)^2 
(\sigma-\sigma_Q)^2 + \lrS{\mu_0 + \mu_1 \sigma} 
\tau + \f{\nu_0}{2} \tau^2.
\qe
The first term with the constants $\sigma_P$ and $\sigma_Q(>\sigma_P)$
induces density phase separation, while the coefficients
$\mu_0$, $\mu_1$, and $\nu_0$ control shear elasticity.
When $\mu_0>0$ and $\mu_1=\nu_0=0$, the form of the shear free
energy agrees with what prescribed by the classical
affine-deformation theory of rubber-elasticity~\cite{FloryBook}.
The coefficient $\mu_1$ accounts for the coupling
between density and shape anisotropy,
and $\nu_0$ is introduced to ensure stability
when $\mu_0$ or $\mu_1$ is negative.
%
\par
Because of the nonlocal nature of elastic distortion,
phase coexistence in a gel crucially depends on 
boundary conditions and geometry, and 
admits no simple and exact analysis like Maxwell construction~\cite{SK}.
Here we restrict ourselves to the configuration of free energy 
minima in the $\sigma-\tau$ plane, which helps
classification of phase behavior and 
drawing an approximate phase diagram. 
For this purpose it is useful to 
define the effective shear modulus,
\eq
\mu = \der{f}{\tau} = \mu_0 + \mu_1 \sigma + \nu_0 \tau.
\qe
Minima of $f$ under the constraint
$\sigma\le\tau$ is controlled by the location of the line $\mu=0$
with respect to the points $P:(\sigma_P,\sigma_P)$
and $Q:(\sigma_Q,\sigma_Q)$ on the $\sigma-\tau$ plane (Fig.1).
There are four possibilities:
(i) isotropic-isotropic (I-I) separation:
$P$ and $Q$ locate the minima if both $\mu|_P$
and $\mu|_Q$ are positive;
(ii) AG-I separation:
if $\mu|_P < 0$ and $\mu|_Q>0$,
the free energy minimum corresponding to
the collapsed phase is on the line $\mu=0$ 
and off the isotropic line $\sigma=\tau$;
(iii) I-AG separation ($\mu|_P >0$, $\mu |_Q <0$):
the swollen phase is accompanied by spontaneous shape anisotropy;
and
(iv)  AG-AG separation ($\mu |_P <0$, $\mu |_Q <0$).
%
\par
The total free energy is the sum of homogeneous
and gradient contributions, and is assumed in the form
\eq
F = \int d\br \biggl[
f(\sigma,\tau) 
+ \f{L}{2}\lrS{\grad \sigma}^2  
+ \f{M}{2}\lrS{\grad \tG}^2  
\biggr].
\qe
While the second term penalizes density gradients
and drives domain coarsening, the last term penalizes 
also gradients of the principal strain axis,
and drives orientational ordering.
To describe the latter, we define a ``nematic'' 
order parameter $Q_{ij} = G_{ij}-G_{kk}\delta_{ij}/2$~\cite{GL}.
%
\par
For the dynamics, we assume the simplest
equation for gels with viscous 
solvents~\cite{SSK,HK},
\eq
\der{\bR}{t} = - \Gamma \fder{F}{\bR},
\label{Rdyn}
\qe
where $\Gamma$ is the inverse of viscosity.
Linearizing with respect to the displacement $\bu=\bR-\br$,
we have the initial growth of the longitudinal mode
$\bu_{\parallel}(\bq)= q^{-2} \bq \bq \cdot \bu(\bq)$ as
\eq
\der{}{t} \bu_{\parallel}(\bq) 
&=& - \Gamma \lrL{V + W + (L+4M) q^2} q^2
\bu_{\parallel}(\bq),
\\
V &=& \lrS{\der{^2f}{\sigma^2} + 2 \der{^2f}{\sigma \D \tau} +
\der{^2f}{\tau^2}} \bigg|_{\sigma=\tau=1}
\nn\\
&=& 2 (\sigma_P+\sigma_Q-2)^2 + 2(\sigma_P-1)(\sigma_Q-1) 
\nn\\ 
&+& 2\mu_1 + \nu_0,
\\
W &=& \mu \big|_{\sigma=\tau=1} 
= \mu_0 + \mu_1+ \nu_0,
\qe
while the transverse mode $\bu_{\perp} = \bu - \bu_{\parallel}$
grows as
\eq
\der{}{t} \bu_{\perp}(\bq) = - \Gamma \lrS{W + 2 M q^2} q^2 \bu_{\perp}(\bq).
\label{growthperp}
\qe
These give the spinodal conditions for density
separation and orientational ordering 
as $V+W<0$ and $W<0$, respectively.
%
\par
To study the nonlinear regime, we implemented
the model on a $256\times256$ square lattice with periodic 
boundary conditions.
The kinetic equation (\ref{Rdyn}) was time-integrated with
the Euler scheme. 
The reference parameters are 
$\sigma_P=0.6$, $\sigma_Q=1.5$, $L=0.2$, $M=0.05$, $\mu_0=-0.11$,
$\mu_1=\nu_0=0.05$, and $\Gamma=0.15$, with the time increment
and grid size set to unity.
For this set of parameters, the initial state is unstable
both to phase separation and orientational ordering
(as $V=-0.23$ and $W=-0.01$).
For any field quantity $A(\br)$, we shall denote 
its spatial averages over the collapsed ($\sigma<1$) 
and swollen ($\sigma>1$) regions by $\{A\}_<$
and $\{A\}_>$, respectively.
%
\par
Figure 2(a) shows evolution
of the network density field $1/\sigma(\br)$.
At an early stage, there appears a foam-like pattern
connected by strands of the collapsed phase.
Then the strands start to undulate.
At each vertex of the foam-like pattern,
the strands fold onto themselves to form a lump
embracing two spots of the swollen phase.
Finally the pattern coarsens in a self-similar
fashion, with small vertices and thin strands dissolving.
Plotted in Fig.2(b) is the distorted lattice mesh
defined by $\bR(m\be_x+ n\be_y)$($m,n$: integers),
and the corresponding ``Schlieren'' texture $Q_{xy}^2$.
Note that each folded structure contains 
a pair of 1/2 defects. 
%
\par
To see the origin of the morphology,
next we study the crossover between AG-I and I-I separation.
The domain boundary becomes less wiggly
as $\mu_0$ is increased, and the effective shear
modulus in the collapsed phase $\{\mu\}_<$
turns positive at $\mu_0=-0.076$; see Fig.3.
For larger values of $\mu_0$, the model reproduces
the foam-like morphology known for I-I separation~\cite{SSK}.
In the bottom of the figure, we plot versus $\mu_0$ 
the principal elongation ratios in the two phases.
In I-I separation, the swollen phase is almost isotropic
($\lambda_1\simeq\lambda_2$), while
the collapsed phase has a sizable strain anisotropy. 
This anisotropy comes from the volume mismatch 
between the two phases. At a flat phase boundary, 
continuity of the network mesh requires elongation 
of the collapsed region parallel to the interface,
which creates elastic tension; see Fig 4(a).
%
\par
In AG-I separation, on the other hand, the AG region
chooses its anisotropy axis along the interface to
reduce the tension. The boundary stress is
completely canceled when the ratio of the spontaneous elongation
matches that of the boundary-induced anisotropy.
This is roughly identified with the case
$\{\lambda_1\}_< = \{\lambda_1\}_>$,
which occurs at $\mu_0=-0.088$.
For larger anisotropy,
or when $\{\lambda_1\}_< > \{\lambda_1\}_>$,
the mismatch stress is canceled by
tilting the strain axis in the AG phase,
as illustrated in Fig.4(b).
At boundaries between interfacial regions
with different tilt directions, however, continuity of 
the network mesh requires a buckling of the interface.
At the same time, torque is created at vertices
where different strands meet, since the tilt is
opposed by the clamping at the center; see Fig 4(c).
This torque causes a spontaneous winding of the structure, 
which is further cut at two points near its center to 
reduce elastic free energy, creating two defects (Fig.4(d)).  
This leads to the patterns as seen in Fig 2.(b).
%
\par
Now we consider the case of I-AG separation by
assuming a negative density-anisotropy coupling.
Shown in Fig.5(a) is the coarsening process
for the parameters $\mu_0=0.05$ and $\mu_1=-0.11$.
Initially a foam-like morphology like in 
I-I separation appears.
As anisotropy in the AG phase develops,
the cell walls break and the swollen domains merge.
Finally the isotropic phase form droplets,
as is the case for liquid-liquid phase separation
with the same volume fraction. 
In Fig.5(b), we illustrate the deformation induced
around an droplet of the isotropic phase. To reduce 
the mismatch at phase boundary, the principal elongation 
axis in the AG phase tends to be perpendicular to the 
interface. Therefore, the deformation in the 
AG phase is necessarily inhomogeneous.
%
\par
The formation of the droplet morphology 
is understood from the fact that an AG
in inhomogeneous deformation states is
liquid-like soft~\cite{lcgel2,Unpub2}.
Internal elastic stresses are greatly reduced 
due to a screening of strain-mediated long-range 
interaction between orientationally correlated regions.
In the present case, it means that stresses 
created at a phase boundary are not transmitted 
far into the AG regions. Thus, the foam-like 
pattern cannot be mechanically supported and 
the morphology is controlled by interfacial tension, 
leading to the liquid-like pattern.
%
\par
To summarize, novel patterns should arise 
in phase-separating gels with density-anisotropy coupling.
To realize the wiggly self-folded morphology, we need the
anisotropy ratio $\lambda_1/\lambda_2$ to be $3\sim4$ in 
the AG phase. It is a realistic value in nematic gels
made up of main-chain liquid-crystalline polymers~\cite{Mainchain}.
Also, negative coupling and 
collapsed-isotropic--swollen-nematic separation 
can be achieved by using nematic solvents~\cite{WW}.
The 2D patterns should be observable in a film 
geometry as used in fabrication of polymer-dispersed liquid crystals, 
which basically are isotropic gels with nematic solvents 
and do show quasi-2D patterns~\cite{PDLC}.
It should be interesting to extend the study to a 3D system.
There the swollen isotropic phase creates an isotropic tension
along a flat phase boundary, while uniaxial deformation of 
the AG phase can cancel it only in one direction; the stress 
in the other direction would cause a spontaneous curvature 
of the boundary. 
On the other hand, the liquid-like morphology
would be qualitatively unchanged, since the soft elasticity 
of the AG phase is due to the presence of bulk Goldstone modes.
%
%
\par
I thank Professor Akira Onuki for valuable discussions. 
This work is supported by 
Grant-in-Aid for Scientific Research from Japan Society 
for the Promotion of Science.
%
%
\vspace{15mm}
\\\null\quad
{\small
*Present address: Department of Physics, 
Tohoku University, Sendai 980-8578, Japan. 
}
\vspace{-27mm}\\
%
%

%
%
\begin{figure}[h]
\noindent
\epsfxsize=240pt \epsffile{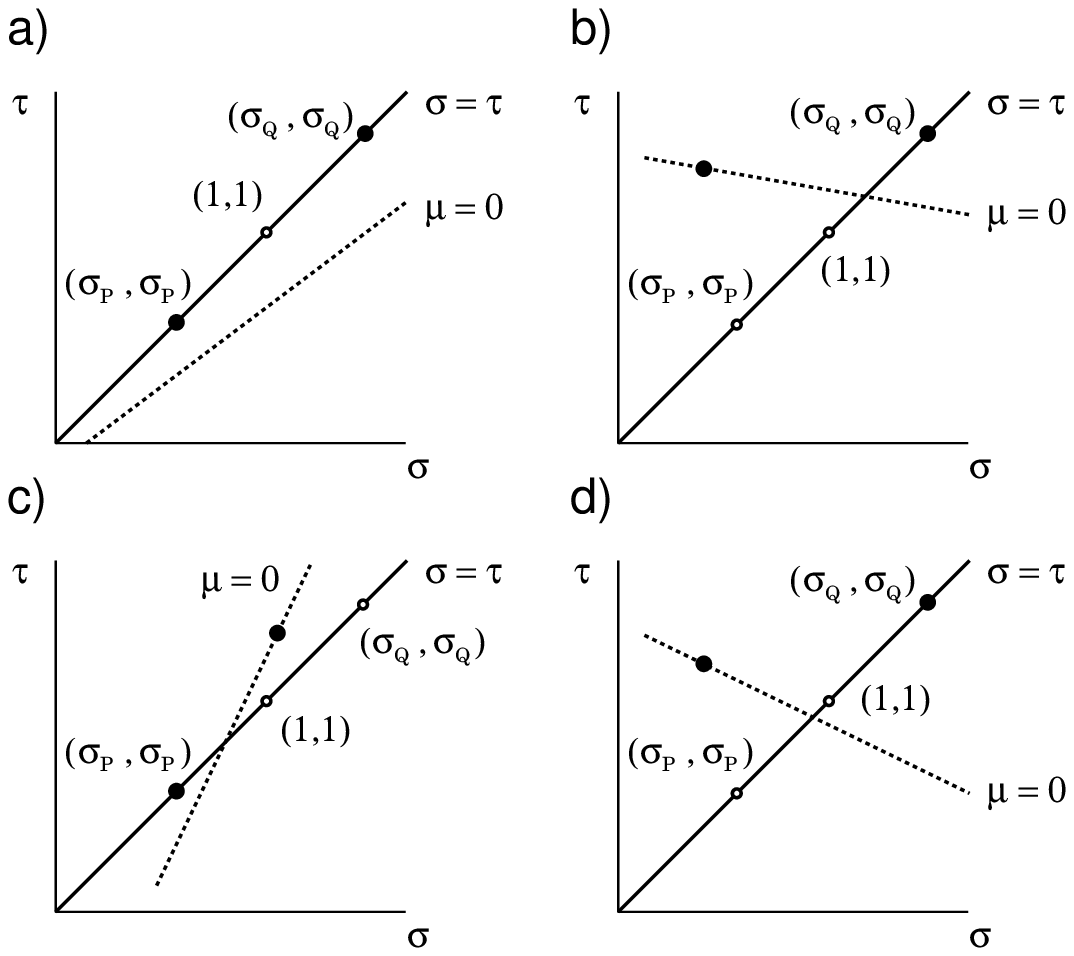}
\mycaption{
\label{fig1}
Schematic configurations of free energy 
minima (marked by $\bullet$). 
(a) I-I separation.  
(b) AG-I separation.
(c) I-AG separation.
(d) AG-I separation with the initial state 
stable against orientational ordering (see Eqs.(7) and (8)).
}
\end{figure}
\begin{figure}[h]
\noindent
\subfigure{\epsfxsize=240pt \epsffile{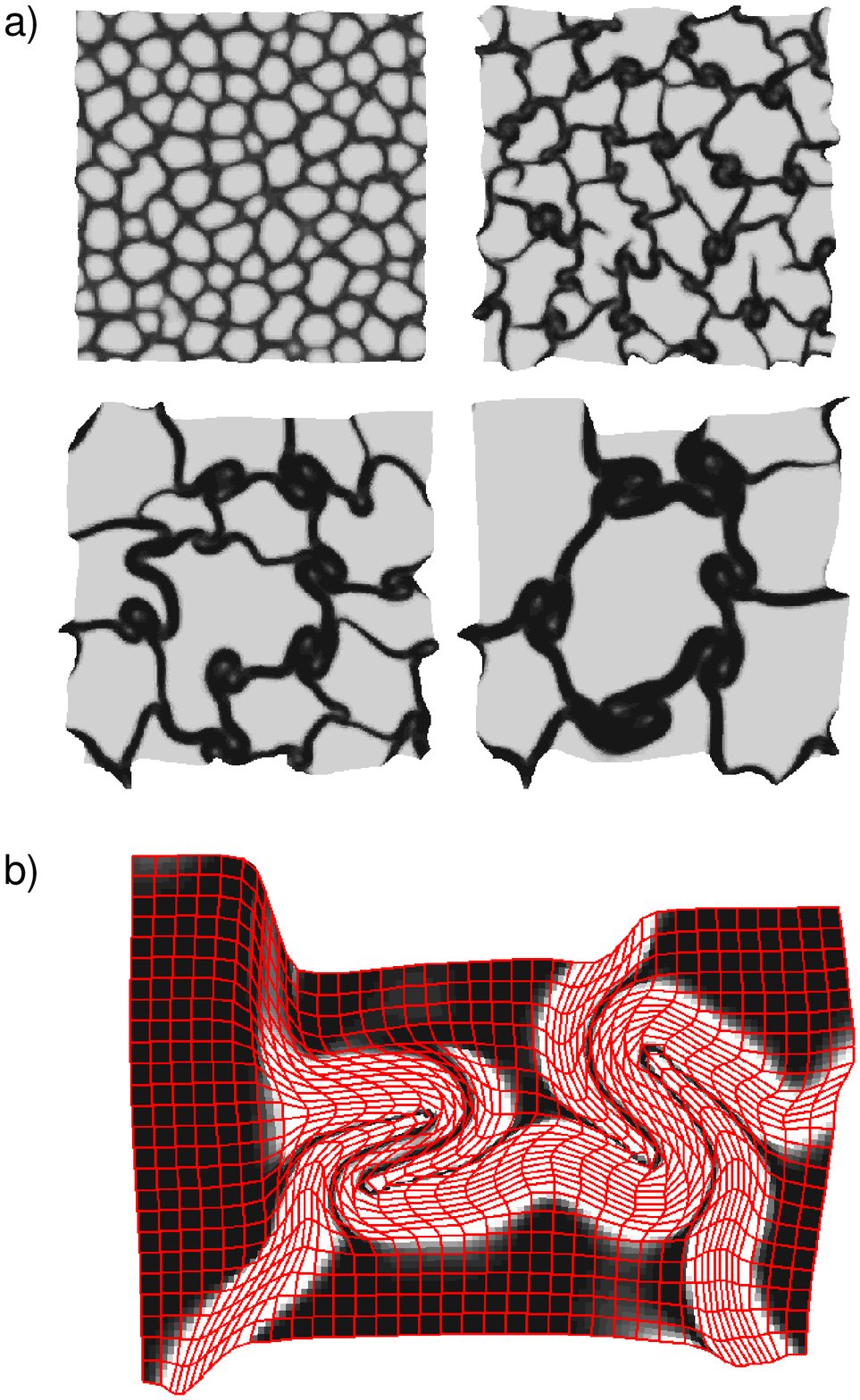}}
\mycaption{
\label{fig2} 
(a) Evolution of the network density field.
Collapsed (AG) regions are plotted in black.
Shown is a $192\times192$ portion of the distorted lattice,
at $t=(1,4,16, \mbox{and } 64) \times 10^4$ from the top left to bottom right.
(b) Distorted lattice mesh (red) and ``Schlieren''
texture $Q_{xy}^2$ (grayscale).
}
\end{figure}
\begin{figure}[h]
\subfigure{\epsfxsize=220pt \epsffile{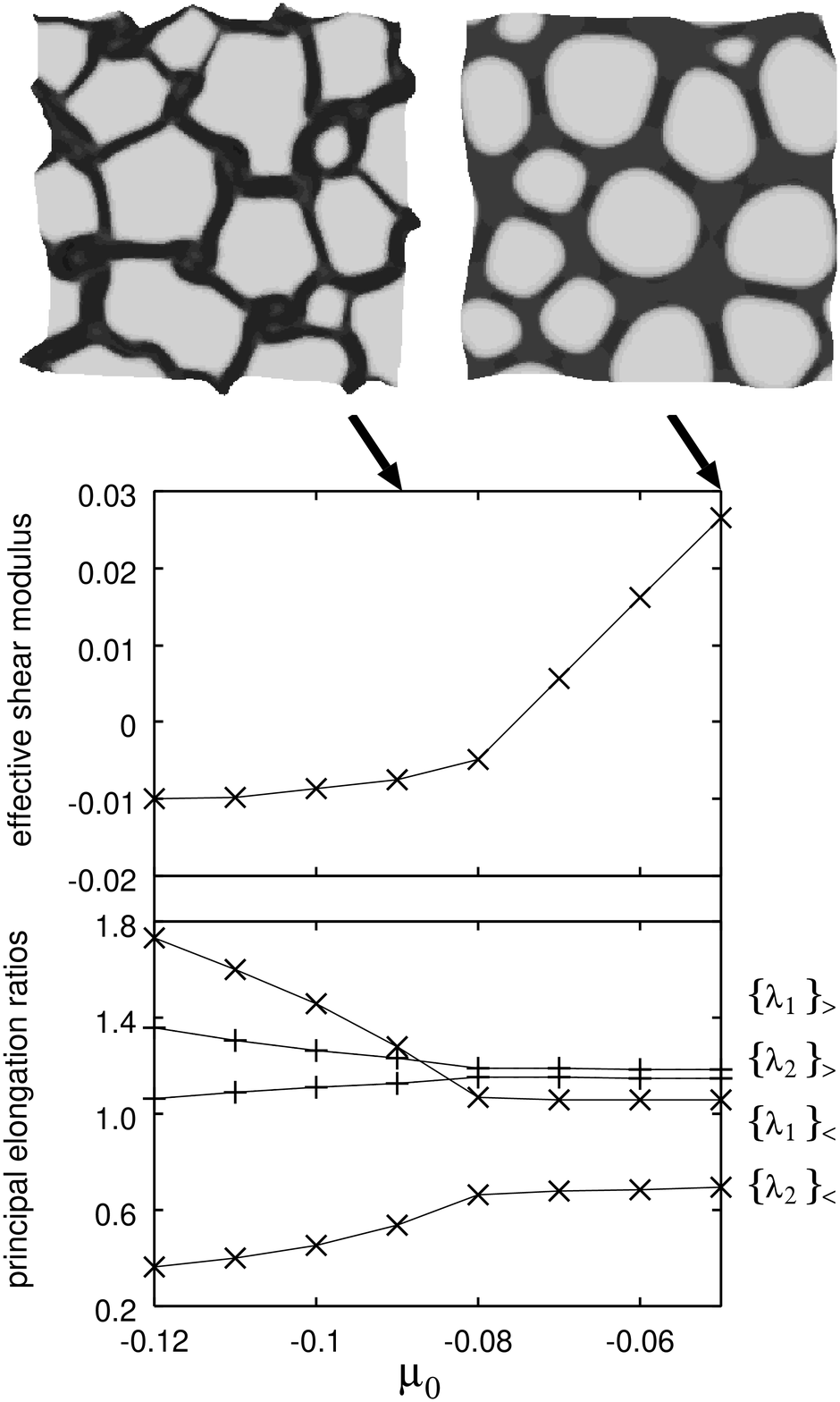}}
\mycaption{
\label{fig3}
Crossover from AG-I to I-I separation.
Top: domain morphologies 
at $\mu_0=-0.09$ and $\mu_0=-0.05$.
Middle: the effective shear modulus averaged 
over collapsed regions, $\{\mu\}_<$.
Bottom: principal elongation ratios averaged over 
collapsed and swollen regions.
All the data are obtained at $t=32\times10^4$.
}
\end{figure}
\begin{figure}[h]
\subfigure{\epsfxsize=220pt \epsffile{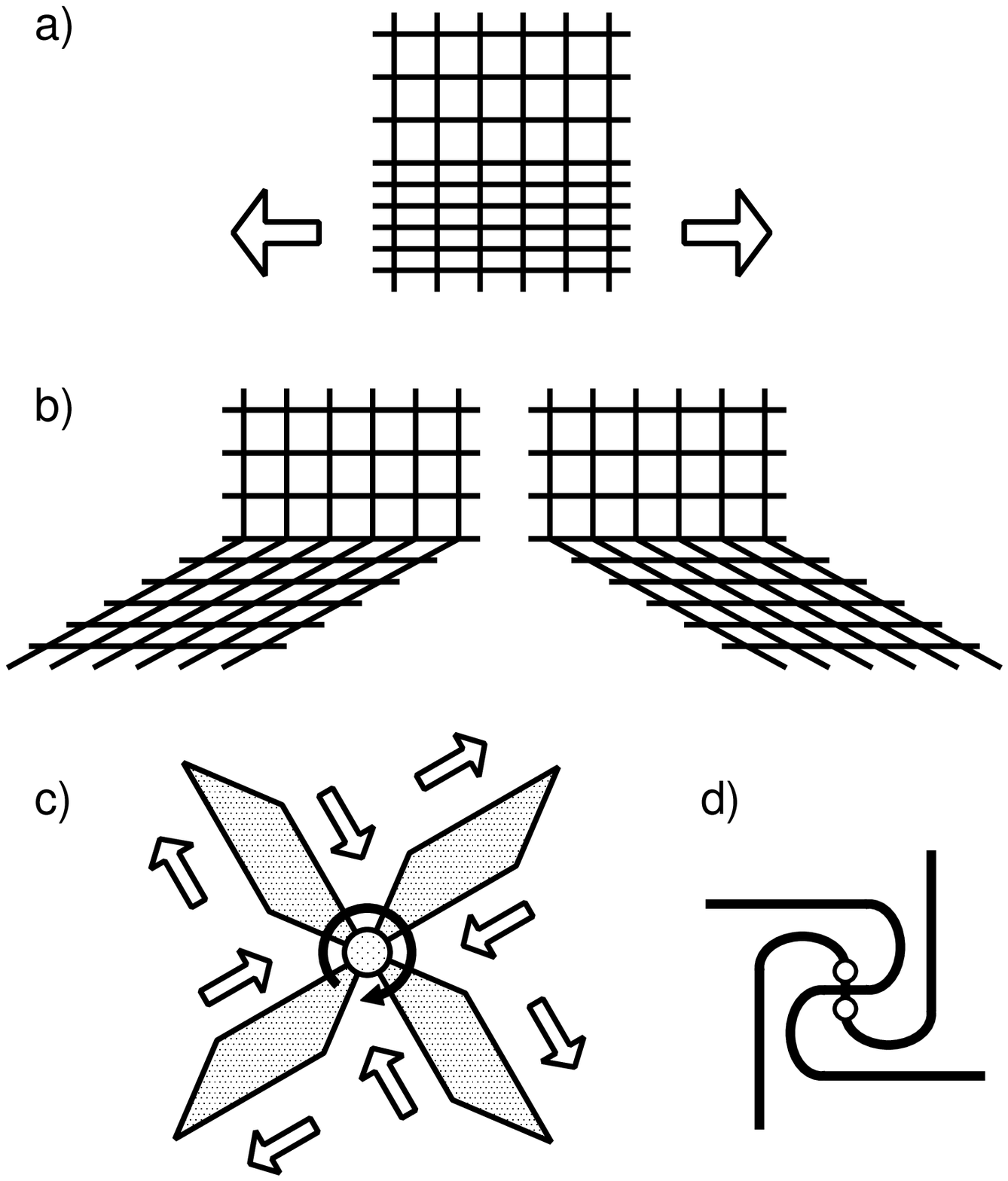}}
\mycaption{
\label{fig4}
Deformation at interfaces.
(a) If the preferred elongation ratio $\lambda_1$
of the collapsed phase is smaller than that
of the swollen phase, mismatch of the network
mesh creates tension along the interface.
(b) In the opposite case,
the mismatch stress is canceled by tilting
the elongation axis in the AG phase.
(c) Torque is created at each vertex,
where tilted strands are clamped.
(d) The cross-shaped structure in (c)
winds up itself and is cut at the
points marked by $\circ$.
}
\end{figure}
\begin{figure}[h]
\noindent
\subfigure{\epsfxsize=240pt \epsffile{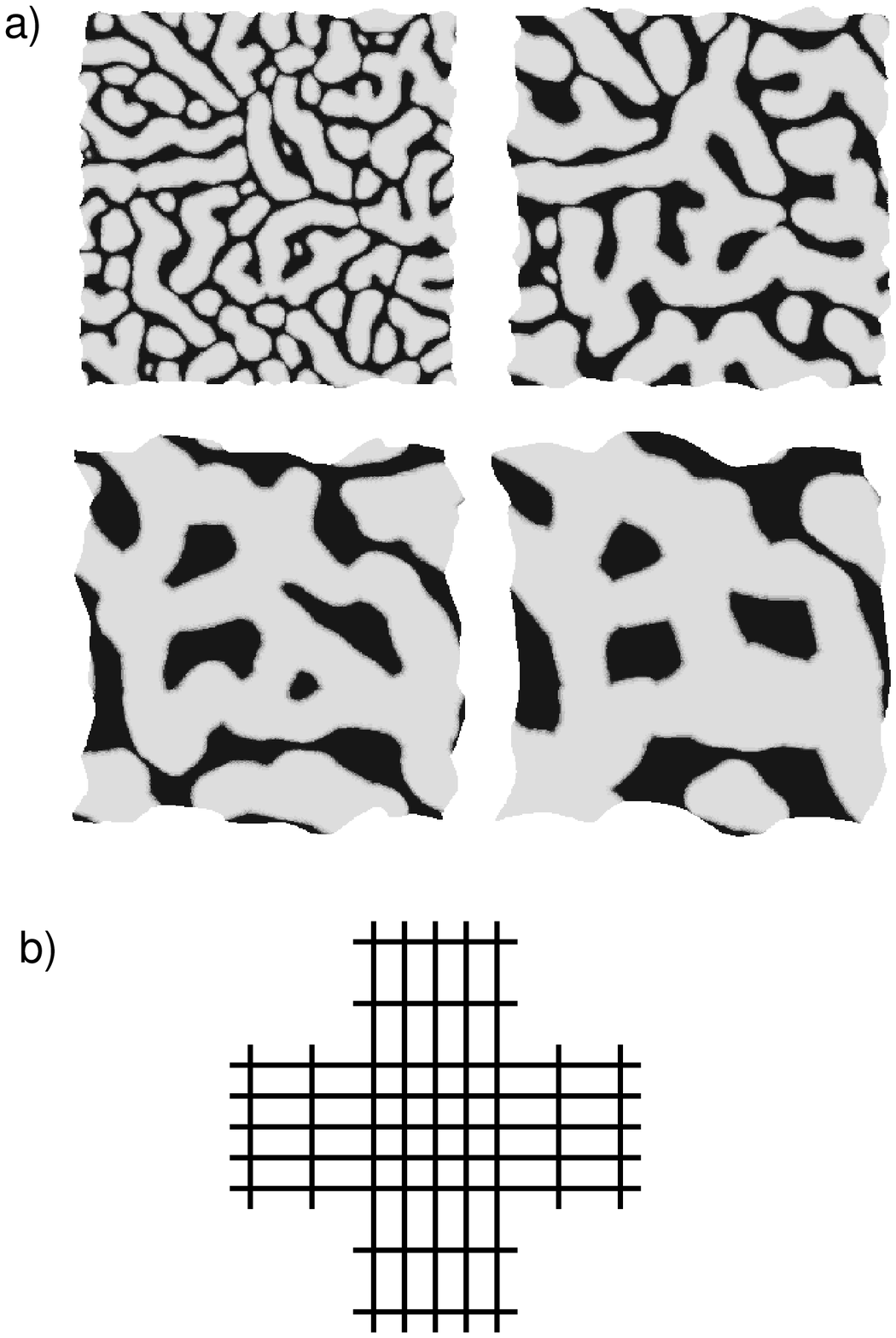}}
\mycaption{
\label{fig5}
(a) Density snapshots for I-AG separation,
at $t=(1,4,16, \mbox{and }64) \times 10^4$ from the top left to bottom right.
(b) Boundary matching induces inhomogeneous anisotropy 
in the AG phase around an isotropic droplet.
}
\end{figure}
\end{multicols} 

\begin{references}
%
\bibitem{TTanaka}
T. Tanaka {\it et al.},
Nature {\bf 325}, 796 (1987); 
E. S. Matsuo and T. Tanaka, Nature {\bf 358}, 482 (1992). 
%
\bibitem{SSK}
K. Sekimoto {\it et al.},
Phys. Rev. A {\bf 39}, 4912 (1989).
%
\bibitem{OnP}
A. Onuki and S. Puri,
Phys. Rev. E {\bf 59}, R1331 (1999).
%
\bibitem{HTanaka}
H. Tanaka, Phys. Rev. Lett. {\bf 71}, 3158 (1993).
%
\bibitem{PDLC}
K. Amundson {\it et al}, 
Phys. Rev. E {\bf 55}, 1646 (1997);
J. B. Nephew {\it et al},
Phys. Rev. Lett. {\bf 80}, 3276 (1998).
%
\bibitem{GL} 
L. Golubovi\'c and T. C. Lubensky, 
Phys. Rev. Lett. {\bf 63}, 1082 (1989).
%
\bibitem{deGennes}
P. G. de Gennes, C. R. Seances Acad. Sci. B {\bf 281}, 101 (1975).
%
\bibitem{Finkelmann}
H. Finkelmann {\it et al.},
Makromol. Chem. Rapid Commun., {\bf 2}, 317 (1981).
%
\bibitem{lcgel1}
N. Uchida and A. Onuki, Europhys. Lett. {\bf 45}, 341 (1999).
%
\bibitem{lcgel2}
N. Uchida, Phys. Rev. E {\bf 60}, R13 (1999); {\bf 62}, 5119 (2000).
%
\bibitem{OM}
Spinodal fluctuation in nematic gels is already discussed by:
P. D. Olmsted and S. T. Milner, Macromolecules {\bf 27}, 6648 (1994).
%
\bibitem{OoP}
Y. Oono and S. Puri, Phys. Rev. Lett. {\bf 58}, 836 (1987).
%
\bibitem{FloryBook}
P. J. Flory, Principles of Polymer Chemistry
(Cornell University, Ithaca, 1953).
%
\bibitem{SK}
K. Sekimoto and K. Kawasaki, 
Physica A {\bf 154}, 384 (1989).
%
\bibitem{HK}
T. Hwa and M. Kardar, Phys. Rev. Lett. {\bf 61}, 106 (1989).
%
\bibitem{Unpub2}
The liquid-like softness of an inhomogeneous AG has been 
reproduced by the present model (modified for a one-phase system):
N. Uchida, unpublished results.
%
\bibitem{Mainchain}
G. H. F. Bergmann {\it et al.},
Macromol. Rapid Commun. {\bf 18}, 353 (1997).
%
\bibitem{WW}
X. J. Wang and M. Warner, Macromol. Theory Simul. {\bf 6}, 37 (1997).
%
\end{references}
\end{document}